\documentclass{sf2a-conf2026}
\usepackage{graphicx}
\usepackage{hyperref}
\usepackage[]{natbib}  
\usepackage{epstopdf}

\def\BibTeX{{\rm B\kern-.05em{\sc i\kern-.025em b}\kern-.08em
    T\kern-.1667em\lower.7ex\hbox{E}\kern-.125emX}}
\bibpunct{(}{)}{;}{a}{}{,}  %%%%%%%%%%%%%  A&A bibliography style
\begin{document}

\TitreGlobal{SF2A 2026}

\title{Limits to research in astrophysics}
% from the perspective of  researchers?

\runningtitle{Limits SF2A-2026 S13}
% 4 pages

\author{F.~Cantalloube}\address{Univ. Grenoble Alpes, CNRS, IPAG, F-38000 Grenoble, France}
\author{C.~Noûs}\address{Laboratoire Cogitamus, \url{https://www.cogitamus.fr/}}
%% Keep this line, even if the page will be settled afterwards.
\setcounter{page}{237}

\maketitle

\begin{abstract}
Fundamental physics faces several intrinsic limitations, such as the finite speed of
light in vacuum, the absolute zero temperature, Heisenberg’s uncertainty principle and so on. At the same time, the material resources necessary to conduct research in astrophysics, such as energy, fresh water, and rare-earth elements, are also limited in both quantity and extraction capabilities, a concept often referred to as \emph{planetary boundaries}. 
In the context of scientific research growing at a frenetic pace, following the greedy quenchless extractivist model of our era, and which is giving rise to an ever-increasing number of large-scale, increasingly ambitious projects: what limits us, individually and collectively, in the practice of our occupation ?
In this conference proceeding, we discuss certain limiting factors we encounter in our daily work, which we need to reflect on collectively in order to move forward, toward a more sustainable and effective research that preserves the well-being of researchers, regardless of their career stage.
\end{abstract}

%% Insert the keywords (to appear in the ADS indexing)
%% Keywords must be separated by a comma

\section{Introduction}
This talk was initially built in the context of an art-science research project (GAES-2025, coordinated by the \emph{Hexagone} in Grenoble and the \emph{Quai des Savoirs} in Toulouse), which topic was "Knowledge of Limits, Limits to Knowledge"  [Savoirs des limites, Limites aux savoirs]. 
Below, I offer a list of personal reflections on obstacles (some major, some minor) that affect the way we currently conduct research. This list is by no means exhaustive or universal: some obstacles may be missing, while others presented as major may seem insignificant to some readers, and vice-versa. The idea behind this perspective is simply to offer suggestions that help us recognize the areas where our current approach to conducting research is not sustainable in the long term.

As I am writing this SF2A proceeding, on Monday 5th of October 2026\footnote{Finishing writing, 800,000 protesters supported the high-school students and their demands on the 17th of October.}, under $26^\circ\mathrm{C}$\footnote{Weather showed a sudden drop by $13^\circ\mathrm{C}$ on the 17th of October.}, under 18 high-school students are being violently repressed by the police, under the order of the current French government. %This government built against the results of the French legislative elections in summer 2024. 
In less than a week, as $400$ high-school students were holding a blockade, there were $1747$ arrests of school-aged pupils who are protesting against the deleterious studying conditions. These protests were violently repressed with disproportionate use of tear-gas, grenades and blast balls. The interior minister declared today that police are granted "no instructions to hold back and unrestricted use of all available intermediate weapons [aucune consigne de retenue et aucune consigne de ne pas utiliser tout l'armement intermédiaire à disposition]"\footnote{https://x.com/BFMTV/status/2106088835113447573}. 
So, before diving into other aspects, let's make it crystal clear in one sentence: as demonstrated by history, one ultimate limit to fundamental research is entering a totalitarian fascist regime. %In France, it is undeniable that in the last few years, totalitarian and fascist regime is on its way.
In a burning world we cannot do research, whatever the origin of the conflagration. %The last few years witnessed it all over the world, thinking that Europe is protected from such fate is a dangerous misconception.

%%---------------------  
\section{A bundle of limits}
%\paragraph{Fundamental limits} 
%Limite spatiale, temporelle, personnelles.

One very obvious fundamental limit, encompassing everything, is the temporal limit. Considering the astronomical temporal scales, astronomical events may take place on timescales longer than humanity's lifetime. On shorter timescales, and for the sake of repeatability needed in most observations, some events over a few centuries might be bypassed. There is also the limit of the observable universe due to the finite speed of light. % that gives access to temporal slice surrounding observatories. 
Another fundamental limit, is the size of the collecting areas of the astronomical observatories, which cannot exceed the size of the Earth for ground-based telescopes (with the example of the Event Horizon Telescope combination) or the stability of laser coupling for space-based interferometric projects. This limit is to be approached with ultimate sensitivity limit such as the traditional photon noise limit for electromagnetic observatories or similar for other astrophysical signals / messengers. At last, astrophysics, relying on huge infrastructures, is particularly conditional to energy access: when it runs out, what and who will determine whether these can still operate.
%without energy, these infrastructures can no longer function.

In the following, the focus will not be on such fundamental limits. To ease the reading, I attempted to sort out the so-called \emph{limits} discussed during the session into three categories: methodological, collective, and personal. Once again, these categories are arbitrary and may overlap with one another. 
%slides link ?

\subsection{Methodological limits}
In this section I purposely ignore any mention of the use of generative artificial intelligence (AI) techniques that are widely debated in this context, especially during the last two years. Past research is feeding AI.

\paragraph{Tree structure and big picture}
The way research is conducted is usually incremental, going deeper and deeper into the comprehension of a \emph{problem} (in its general term). It also means that often \emph{solving} one aspect opens many questions, often more complex. This leads to overspecialization, and its corollary: losing sight of the big picture. One example in exoplanet imaging: about 20 years back, a handful of researchers worked together on both the instrument design, the data collection on-site, and the analysis and interpretation of the data. Nowadays, exoplanet imaging expertise is crucially chopped down: some researchers are expert of a certain type of subsystem of a key ingredient of the instrument. Expertise in data interpretation is also fragmented, for example atmospheric and dynamic model fitting in exoplanet science, also depending on the detection method.
%Expertise in data interpretation is also chopped, for example atmospheric and dynamic model fitting in exoplanet science, itself different according to the detection method.
%\emph{For instance, not only into instrumentalist, but into e.g. expert in adaptive optics and even deeper into expert of one specific type of wavefront sensor. Same for the data interpretation, with e.g. expertise in atmosphere fitting, and expertise in dynamic fitting being often separated due to the inherent complexity of both expertise.}%too complex

\paragraph{Data growth}
One elephant in the room is the huge amount of data that is nowadays collected. Not only the actual scientific target data, but also the ancillary data such as telemetry or any environmental data that may be used to refine the data analysis. Archival data are more and more massive, while they need to be properly converted, transferred and stored (either raw, processed or by-products). In the case of ESO, statistics on the data collected, exploited and published \citep{esodata} can be found here \url{https://www.eso.org/sci/php/libraries/pubstats/} and strategies for the future ELT are being formalized. Future observatories such as the SKA observatory will realistically not run at 100\% observing duty cycle because it simply cannot save everything it measures. Later in the process of exploiting the data, current publishing policy, linked to ethical science practice, require to publicly provide with the analysis tools and the data along with the scientific publications. This comes to a question rising during the last few years: the human resources needed to process, ingest, understand and keep track of the data are currently not sufficient in face of the growing number of data incoming and their complexity. Also, catching up with all these discoveries becomes challenging. %and working on the data starts to be difficult. %Un corrollaire: seules les IA seront capables de se tenir a jour des decouvertes %SKA cannot run at 100% duty cycle

\paragraph{Model vs. Data} The data collected are more and more precise, but often the physical models are not ready to perform the fitting despite their greater degrees of complexity with added physical and chemical effects. This inherent limit is well illustrated by exoplanet atmospheric data and models: the JWST spectrographs offer an extravagant spectral resolving power, while existing models do not account for all the details via reliable molecular data and atmospheric physics \citep{petrus2024jwst}. A similar effect occurs with higher spatial resolution, with the example of the Solar imaging where small-scale features are observed but their interpretation is far from obvious and requires subsequent work to reach such level of comprehension \citep{schmidt2025}. 
%JWST's unprecedented spectroscopic precision is exposing limitations in current theoretical models: systematic uncertainties in atmospheric structure, cloud physics and molecular opacities can prevent models from reproducing the observations reliably, limiting the physical parameters that can be inferred from the data.

\paragraph{Loss of expertise}
The introduction of new technologies, which are not necessarily synonymous with progress, may provoke a loss of expertise or know-how. As an old-fashion example, detectors are now quantifying the weather conditions and weather forecast are heavily used to prepare for the observing night. Astronomers used to spend more time outside and analyzing weather patterns in preparation for the night, optimizing the schedule based on their experience. Another example would be the complexity of instruments: similarly to a contemporary car engine that has so many parts and electronic components that most people cannot know how to fix it. In addition, for detectors, or other sensors, military applications are never too far and some technologies are patented/classified, preventing public scientists from understanding their inherent functioning. 
%AI stop publishing+the openAI maths dump. How soon they do before the whole ESO archive ? 

\subsection{Collective limits}
\paragraph{Memory}
The publication language of most science, and especially astrophysics and optics, used to be French, German and English until the 80s, when it turned to english-only. Automatically some old publications are no longer accessible to the wider community of astrophysicists. One example is the (re-)discovery of the \emph{low wind effect} in high-contrast imaging: this effect, due to cold telescope spider, was noticed, analyzed, interpreted and published in 1949 \citep{couder1949lwe}, in French. When in 2014 this same effect was unveiled on the ESO Very Large Telescope \citep{sauvage2015lwe}, the community was fortunate to have a senior French researcher from Nice who dug out the 1949 publication. When such generation retires, and with the lack of renewed dedicated archivist into observatories/research institute libraries, it will be more and more difficult to keep track of former studies. 

\paragraph{Transmission}
Knowledge is transmitted mainly through studies and PhD at university. It therefore depends on the teacher's knowledge and willingness to add details to a specific topic. As the curriculum are becoming more and more complete and complex, following the natural advances of knowledge, it is more and more tricky to add all we know about a given subject and some pieces of information might be lost on the way. Outreach activities, whatever the media, also participate to the transmission of knowledge and ideas to a general audience. This is another place where knowledge can be trimmed, reduced to the essentials, or lead to misunderstandings.

\paragraph{Bibliography and archives}
On a similar note, there are an exponential rise of scientific publications, even more so for the last few years in the wake of generative AI. For instance, the three main monthly journals: \emph{Astronomy \& Astrophysics} (A\&A), \emph{The Astrophysical Journal} (ApJ), and  \emph{Monthly Notices of the Royal Astronomical Society} (MNRAS) together published 10,454 papers during 2025 (compared to 8000 in 2015 and 5117 in 2005). For both A\&A and ApJ, an increase by 40\% of publications is observed between 2023 and 2024 (going from about 2500 to 3500 papers). Conversely, A\&A stopped hard-over printing of the publications 10 years ago \citep{aanda_online} and scientific papers are published solely online. Such growth is very difficult to follow, starting PhD candidate must focus on their specific topic, while letting out the big picture, in favor of overspecialization. Senior researchers accumulated knowledge that might not be transmitted easily to newer generation of researchers who are exhorted to produce results fast, publish and acquire the overall context and main issues. Also to be published, these papers must be peer-reviewed, increasing the demand for expert reviewers. %\footnote{https://www.bioxbio.com/subject/astronomy}. %created in 1969,

\paragraph{Centralization of Knowledge: prestige and elitism}
As written previously, the main language of astrophysics is nowadays English, which means that westernized countries do have an advantage into the accessibility of these publications. 
%A\&A, created in 1969, used to accept French, German and English language until the 80's (mainly due to difficulties in finding reviewers). 
On a more local note, at the scale of a country like France, the research centres and university are often centralized in and around the capital city Paris, leading to elitism, taking roots in the schools people are able attend, often resting on their sociological background, especially on the wealth, cultural and affective capital. In addition, researchers (who are in charge of hiring/evaluating their colleagues), tend to support people they feel close together, who went to the same education system, and share the same values. This structurally reproduces existing institutional hierarchies and therefore contributes to the reproduction of similar sociological profiles across generations of researchers, favoring researchers trained in prestigious institutions \citep[the so-called homophily effect;][]{berne2020inequalities, wapman2022quantifying, clauset2015systematic}. These hierarchies also influence research priorities and the spread of scientific ideas as mentioned above. Similarly, scientific papers, observation and grant proposals acceptance are also affected by such effect (Example of the gender disparity in success rates for Hubble Space Telescope proposals \url{https://esahubble.org/forscientists/announcements/sci18003/}).
%When looking at the distribution of the IAU members, and their background, the centralization and elitism is quite visible. 
%academic hiring and career advancement can reproduce existing institutional hierarchies and social networks, favouring researchers trained in the same prestigious institutions or in closely connected academic environments. The evidence is particularly relevant to physics and other STEM fields.
%the structural reproduction of the same institutional and educational backgrounds within academia. Social reproduction and cumulative advantage (resistant to changes)

\paragraph{Individualization of research}
Because elitism rests on the idea that scientific progress is driven by exceptional individuals, researchers are encouraged to cultivate their scientific identity as unique, remarkable, and necessary to the community. This is amplified by the way researchers are recruited, as candidates must demonstrate and perform how brilliant they are, compared to a pool of researchers competing for a limited number of often precarious positions. Even after securing a permanent position, the competition carries on through career advancements and grant applications, where success is assessed through quantitative scientific-related criteria such as bibliometry (largely debated), by-passing researcher's ability to teach, mentor or supervise. In the current system, an important discovery is usually let on one single person's publication, although a whole team has worked towards it, and many factors made the discovery possible. This engages disputes, jealousy, frustration and resentment among researchers. 
%Because elitism is itself based on the notion of individuals making the change, researcher are pushed to become obsessed by what makes their scientific identity unique and exceptional (and necessary to the community). This is magnified by the way researchers are recruited, as candidates must demonstrate how brilliant and exceptionnal they are, compared to the pool of precarious researchers in the line. 
Nobel prizes illustrate such effect: one prize is given to one person (now two to three people), regardless of the personality and management skills of the recipient. %And the consequences are such prize is huge on the individual, who is for instance asked about things way outside their field of scientific expertise, although listened to (the so-called halo effect). \footnote{The question of Nobel prizes deserves a whole paper by itself; a list of effects can be found on the dedicated Wikipedia page.}

\subsection{Personal limits}

\paragraph{Exchanging and sharing}
The way researchers communicate have been evolving substantially, from manuscript letters, telegrams and circulars to emails appearing in the 2010s. In the 2020s, other communication platform for instant messaging and document sharing started to be widely used (such as Slack, PicaTeam, Discord, and Mattermost). On top social networks also appeared to exchange information and communicate (such as Mastodont, Bluesky, and Facebook), some even specialized in work networking (such as LinkedIn and ResearchGate). At last, videoconferencing software also took a massive boost in being more and more commonly used, especially during and after the covid-19 pandemics. Because research is based on actors all over the globe, the influx of information is constant. Note that these tools are often link with internal calendars or data repository (which have also multiplied). One example is how observing proposals used to be sent by post-mail (with the stamp being key for the deadline), while now the proposal deposits are online with the same deadline all across the globe. These tools, which were supposed to facilitate our internal communications, have instead, encouraged us to take on more projects \citep[following the rebound effect][]{hennebelle2026} and as a result, overwhelmed us with information and tasks. Now that most of our research results have been digitized, what would happen if, due to energy shortages or any sort of obsolescence, we could no longer access all this numerical data?
%Now each researcher has an average of four email box used for work-related activities. 

\paragraph{Over-inducement}
The number of prizes, awards and grant types has also substantially increased on various level (universities, doctoral schools, societies, scientific journals etc.) and on various topic (diversity, outreach activities, breakthrough discoveries). It takes time and energy to apply and review such awards, affecting self-confidence and humility via a deleterious comparison to peers. This enhances individualism and competition.

\paragraph{Precarity and job insecurity}
Most academic results are produced by early career researchers (ECR) who, in theory, can focus on a specific topic without being disturbed/distracted by too many academic organizing tasks or teaching duties. However, the pressure for publications, job, grant and prize applications, pushes working beyond the contracted hours \citep[average 45~hours a week,][]{frei2022}, while accepting living conditions that are not sustainable and may provoke long-term injuries. The injunction to a mobile international academic career, not only weakens emotional support through relationships and nurturing hobbies, but also affects family plans and friendship, on top of the difficulties due to settling in a new country such as financial difficulties, work-environment shift fatigue and administrative tasks \citep[e.~g.][]{schaer2022mobility}. 
Damages may remain even after making it through to job stability.
Knowledge is transmitted, acquired, extended but most ECR end-up quitting academia \citep[estimations are not exact, but it is more than 50\%,][]{naddaf2024} in a context of enhanced inequalities. 
%, also in order to 'remain competitive' throughout ones career, this pace might be kept for years depending at different level. <- osef c'est leur choix si ils se ruinent tant pis pr eux..., and recognition

\paragraph{Fragmentation and isolation}
Overspecialization fragments research topics, often inducing the feeling of intellectual isolation (to the exception of group leader positions). This can significantly alter mental health, mainly via lack of belonging, loneliness and loss of meaning, while remaining in a very competitive work environment \citep{nicholls2022}. This effect is amplified for under-represented researchers, suffering from discrimination, and often reject \citep[e.g.][]{scherr2020,lewis2016}. The pressure to produce original work also makes one feel that difficulties are personal failures rather than systemic of academia. 
%Intellectual isolation: working on a narrow or highly specialized problem that few people can understand or discuss with you
%when feedback is scarce, or when few people share their scientific interests
%This concerns uncertainty, the pressure to produce original work, competition, precarious career paths, and the sense that one's difficulties are personal failures rather than shared features of research.
%loneliness is a subjective feeling, whereas isolation can describe a person's social or professional circumstances

%%---------------------  
\section{Conclusion and other thoughts}
%Le capitalisme c le manager toxique de nos lifes
%Toutes ces limites, qui chacunes meriteraient leur propre papier, ont en commun qu'elles affectent au dela de la valeur marchande des fondations de notre societe capitaliste, qui pousse a la performance et a la competition dans une temporalite tres restrictive. Continuer a cette vitesse et avec ce  rythme et selon de telles regles n'est pas soutenable.
%Notamment, la sante mentale et la creativite , sans lesquelles les humains, et donc la recherche ne peux s'epanouir. 
All of these limitations, each of which would deserve its own paper, share the common feature that they affect more than just the market value of the foundations of our capitalist society, which drives toward performance and relentless competition within a very restrictive time range. Continuing at such pace, and under such conditions, is not sustainable on the long term. In particular, it affects mental health and creativity, without which humans, and therefore research, cannot flourish. 
The following quote from C.~Nouvian (founder of the BLOOM organization, for the preservation of marine ecosystems, since 2004) speaks for itself: "We destroy what we do not know yet" [On détruit ce qu'on ne connaît pas encore].

%Mais vraiment, comment on avance, comment on ose faire de la recherche dans un monde qui se delite, tant du point de vue de la biosphere que des decisions politiques incitant a la guerre et a la resignation, comment on affronte notre impuissance, nos dissonances et l'alignement avec nos quotidiens. 
Three key questions arise: 
Does the creation of knowledge exempt us, as a collective, from acting, or does it compel us to act? 
How do we dare carrying-on our short-term projects in a world that is falling apart, both in terms of the biosphere and political decisions that push towards war and resignation ? 
How do we have to deal with our own powerlessness, internal conflicts, and how we bring our lives into alignment with our daily lives ?

During outreach talks on these topics, I often conclude with the following sentence "Exploring is refusing to die as humankind" [Explorer c'est refuser de mourir en tant qu'humanité]. %In the current society, the decision for making huge telescopes that may not be run bc of lack of energy

%Ben: you have to mention trump here. Can critique false idea Europe is not going in the same direction.

\begin{acknowledgements}
Many thanks to the SF2A council for making this unique peer-to-peer discussion possible. Thank you to the \emph{SF2A penguin team} and B.~Pope for their support throughout the years on such topics, and for their proofreading and suggestions on the present paper. A wide thank to L\'ea Deshusses from the Hexagone for inviting me to the GAES-2026 and igniting the discussion presented here.
\end{acknowledgements}

%Hyperlinks can be introduced as follows: \url{http://www.sf2a.eu/}.
\bibliographystyle{aa}  
\bibliography{Cantalloube_S13}

\end{document}